# Anisotropic Carrier Transport in *n*-Doped 6*H*-SiC

**R. T. Ferracioli[a], C. G. Rodrigues[a, *], and R. Luzzi[b]**

[a] *School of Exact Sciences and Computing, Pontifical Catholic University, Goiás, Brazil*

[b] *Condensed Matter Physics Department, Institute of Physics Gleb Wataghin, Campinas, Brazil*

**e-mail: cloves@pucgoias.edu.br*

**Abstract**—In this paper, a study is presented on the charge transport in *n*-type doped semiconductor 6*H*-SiC (in both transient and steady state) using a nonequilibrium quantum kinetic theory derived from the method of nonequilibrium statistical operator (NSO), which furnishes a clear description of the irreversible phenomena that occur in the evolution of the analyzed system. We obtain theoretically the dependence on the electric field (applied in the orientation perpendicular or parallel to the *c*-axis) of the basic macrovariables: the "electron drift velocity" and the "nonequilibrium temperature." The "peak points" in time evolution of this macrovariables are derived and analyzed.

## 1. INTRODUCTION

The silicon carbide SiC is a wide-bandgap semiconductor with high saturated carrier drift velocity, small dielectric constant, high thermal conductivity, being an attractive semiconductor for high power device applications [1–5]. Silicon carbide can form in many distinct crystal structures (know as polytypes), with some of the most common being 3*C*-SiC (cubic), 4*H*-SiC (hexagonal), and 6*H*-SiC (hexagonal) [2]. These materials, commonly used for the fabrication of devices, have anisotropic transport properties, because the electron effective masses differ notably depending on orientation along the basal plane perpendicular to the *c*-axis (which we labeled $m^*_{e\perp}$) and that parallel to the *c*-axis (which we labeled $m^*_{e\parallel}$). We draw attention to the fact the degree and characteristics anisotropy are different for each polytype [2].

The optical and transport properties of semiconductors have been studied by using nonequilibrium Green's functions techniques, Monte Carlo simulation, balance equation, etc. In this paper, we use the method of nonequilibrium statistical operator (NSO) [6–15]. The NSO invented by Zubarev is practical and efficient in the study of the transport and optical properties of semiconductors [16–21]. More specifically, we use a Nonequilibrium quantum kinetic theory [22] derived from NSO.

In this work, NSO has been used for the study of bulk nonlinear charge transport in n-type doped semiconductor 6*H*-SiC when the transport direction is along the *c*-axis, or when the transport direction is in the plane perpendicular to it. We obtain theoretically the drift velocity of electrons, $v_e(t)$, and the nonequilibrium temperature of electrons, $T^*_e(t)$, in two regimes: the transient state and steady state. The dependence of these two macrovariables, $v_e(t)$ and $T^*_e(t)$, on the electric field $\mathscr{E}$ applied in the orientation parallel (which we labeled $\mathscr{E}_\parallel$) or perpendicular (which we labeled $\mathscr{E}_\perp$) to the *c*-axis is derived and analyzed. We emphasize that a good knowledge of the transport properties is required to develop high performance electronic devices.

## 2. THEORETICAL BACKGROUND

The Hamiltonian of the *n*-doped semiconductor is taken as composed of: the energy of the free phonons, the energy of the free electrons, the interaction of the electrons with the constant electric field, the interaction of the electrons with the impurities, the electron–phonon interaction, the anharmonic interaction of the LO-phonons with AC-phonons, the interaction of the AC-phonons with the thermal bath (external thermal reservoir at temperature $T_0$).

The nonequilibrium thermodynamic state of the system is very well characterized by the set of *basic variables*:

$$\{E_e(t), N, \mathbf{P}_e(t), E_{\rm LO}(t), E_{\rm AC}\}, \quad (1)$$

that is the energy of the electrons $E_e(t)$; the number $N$ of the electrons; the linear momentum $\mathbf{P}_e(t)$ of the electrons; the energy of the LO-phonons $E_{\rm LO}(t)$; and the energy of the AC-phonons $E_{\rm AC}$.

The *nonequilibrium thermodynamic variables* associated with variables of set (1) are [9–11, 23]

$$\{\beta_e^*(t), -\beta_e^*(t)\mu_e^*(t), -\beta_e^*(t)\mathbf{v}_e(t), \beta_{LO}^*(t), \beta_{AC}^*(t)\}, \quad (2)$$

where $\mathbf{v}_e(t)$ is the drift velocity of the electrons, $\mu_e^*(t)$ is the quasi-chemical potential and

$$\beta_e^*(t) = \frac{1}{k_B T_e^*(t)}, \quad (3a)$$

$$\beta_{LO}^*(t) = \frac{1}{k_B T_{LO}^*(t)}, \quad (3b)$$

$$\beta_{AC}^*(t) = \frac{1}{k_B T_{AC}^*(t)}, \quad (3c)$$

where $k_B$ is Boltzmann constant, $T_e^*(t)$ is the nonequilibrium temperature of the electrons, $T_{LO}^*(t)$ is the nonequilibrium temperature of the LO-phonons, and $T_{AC}^*(t)$ is the nonequilibrium temperature of the AC-phonons [23–25].

By using the Nonlinear Quantum Kinetic Theory based on the NSO, we obtain the equations of evolution for the basic macrovariables [22]:

$$\frac{d\mathbf{P}_e(t)}{dt} = -nVe\mathscr{E} + \mathbf{J}_{\mathbf{P}_{ph}}(t) + \mathbf{J}_{\mathbf{P}_{imp}}(t), \quad (4)$$

$$\frac{dE_e(t)}{dt} = -\frac{e}{m_e^*}\mathscr{E}\cdot\mathbf{P}_e(t) + J_{E,ph}(t), \quad (5)$$

$$\frac{dE_{LO}}{dt} = J_{LO}(t) - J_{LO,an}(t), \quad (6)$$

$$\frac{dE_{AC}(t)}{dt} = J_{AC}(t) + J_{LO,an}(t) - J_{AC,dif}(t), \quad (7)$$

where $n$ is the concentration of electrons fixed by doping (and consequently $dn/dt = 0$); $E_e(t)$ is the energy of the electrons; $\mathbf{P}_e(t)$ is the linear momentum of the electrons; $E_{LO}(t)$ is the energy of the longitudinal optical phonons that interact with the electrons via optical deformation potential and Fröhlich potential; $E_{AC}(t)$ is the energy of the acoustic phonons that interact with the electrons via acoustic deformation potential; $\mathscr{E}$ is the constant electric field applied perpendicular ($\mathscr{E}_\perp$) or parallel ($\mathscr{E}_\parallel$) to the $c$-axis direction. Moreover, in Eqs. (4)–(7), $m_e^* = m_{e\parallel}^*$ when the constant electric field is applied parallel to the $c$-axis, or $m_e^* = m_{e\perp}^*$ when the constant electric field is applied perpendicular to the $c$-axis. These values and other parameters for the study of transport phenomena in 6$H$-SiC are shown in Section 3.

In Eq. (4), the first right-hand term ($-nVe\mathscr{E}$) is the driving force created by the constant applied electric field. The second term, $\mathbf{J}_{\mathbf{P}_{ph}}(t)$, is the rate of electron momentum transfer due to interaction with the LO and AC phonons. The third term, $\mathbf{J}_{\mathbf{P}_{imp}}(t)$, is the rate of electron momentum transfer due to interaction with the ionized impurities [26].

In Eq. (5), the first right-hand term ($-e\mathscr{E}\cdot\mathbf{P}_e(t)/m_e^*$) is the rate of energy transferred from the constant applied electric field to the electrons, and the second term, $J_{E,\,ph}(t)$, is the transfer of the energy of the electrons to the LO and AC phonons.

In Eqs. (6) and (7), the first right-hand term accounts for the rate of change of the energy of the LO or AO phonons, respectively, due to interaction with the electrons, that is, the gain of energy transferred to the phonons from the hot electrons. Like this, the sum of contributions $J_{LO}(t)$ and $J_{AC}(t)$ is equal to the last term in Eq. (5), $J_{E,\,ph}(t)$, but with change of sign. The second term in Eq. (6), $J_{LO,\,an}(t)$, is the rate of transfer of energy from the LO-phonons to the AC-phonons via anharmonic interaction. We noticed that the therm $J_{LO,\,an}(t)$ is the same, however with different sign in Eqs. (6) and (7). Shutting down our analysis of Eqs. (4) to (7), we noticed that the last term in Eq. (7), $J_{AC,\,dif}(t)$, is the diffusion of heat from the AC-phonons to the thermal bath at temperature $T_0$.

We notice that the detailed expressions for the collision operators: $\mathbf{J}_\mathbf{P}(t)$, $J_{E,\,ph}(t)$, $J_{LO}(t)$, $J_{LO,\,an}(t)$, $J_{AC}(t)$, and $J_{AC,\,dif}(t)$ are given in [22].

We emphasize that to close the system of equations of evolution we must establish the relationship between the *basic variables*, Eq. (1), and the *nonequilibrium thermodynamic variables*, Eq. (2). These relationships are:

$$E_e(t) = \sum_{\mathbf{k}} \epsilon_{\mathbf{k}} f_{\mathbf{k}}(t) = nV\left[\frac{3}{2}k_B T_e^*(t) + \frac{1}{2}m_e^* v_e(t)^2\right], \quad (8)$$

$$\mathbf{P}_e(t) = \sum_{\mathbf{k}} \hbar\mathbf{k} f_{\mathbf{k}}(t) = nVm_e^*\mathbf{v}_e(t), \quad (9)$$

$$E_{LO}(t) = \sum_{\mathbf{q}} \hbar\omega_{LO}\nu_{LO}(t), \quad (10)$$

$$E_{AC}(t) = \sum_{\mathbf{q}} \hbar\omega_{AC}\nu_{AC}(t), \quad (11)$$

which are known as "*nonequilibrium thermodynamic equations of state*" [9–11, 23].

In a closed calculation, by using the NESEF, we obtain for phonon populations

$$\nu_{LO}(t) = \frac{1}{e^{\beta_{LO}^*(t)\hbar\omega_{LO}} - 1}, \quad (12a)$$

$$\nu_{AC}(t) = \frac{1}{e^{\beta_{AC}^*(t)\hbar\omega_{AC}} - 1}, \quad (12b)$$

and for electron populations (in the nondegenerate state)

$$f_{\mathbf{k}}(t) = \exp\{-\beta_e^*(t)[\epsilon_{\mathbf{k}} - \hbar\mathbf{k}\cdot\mathbf{v}_e(t) - \mu_e^*(t)]\}, \quad (13)$$

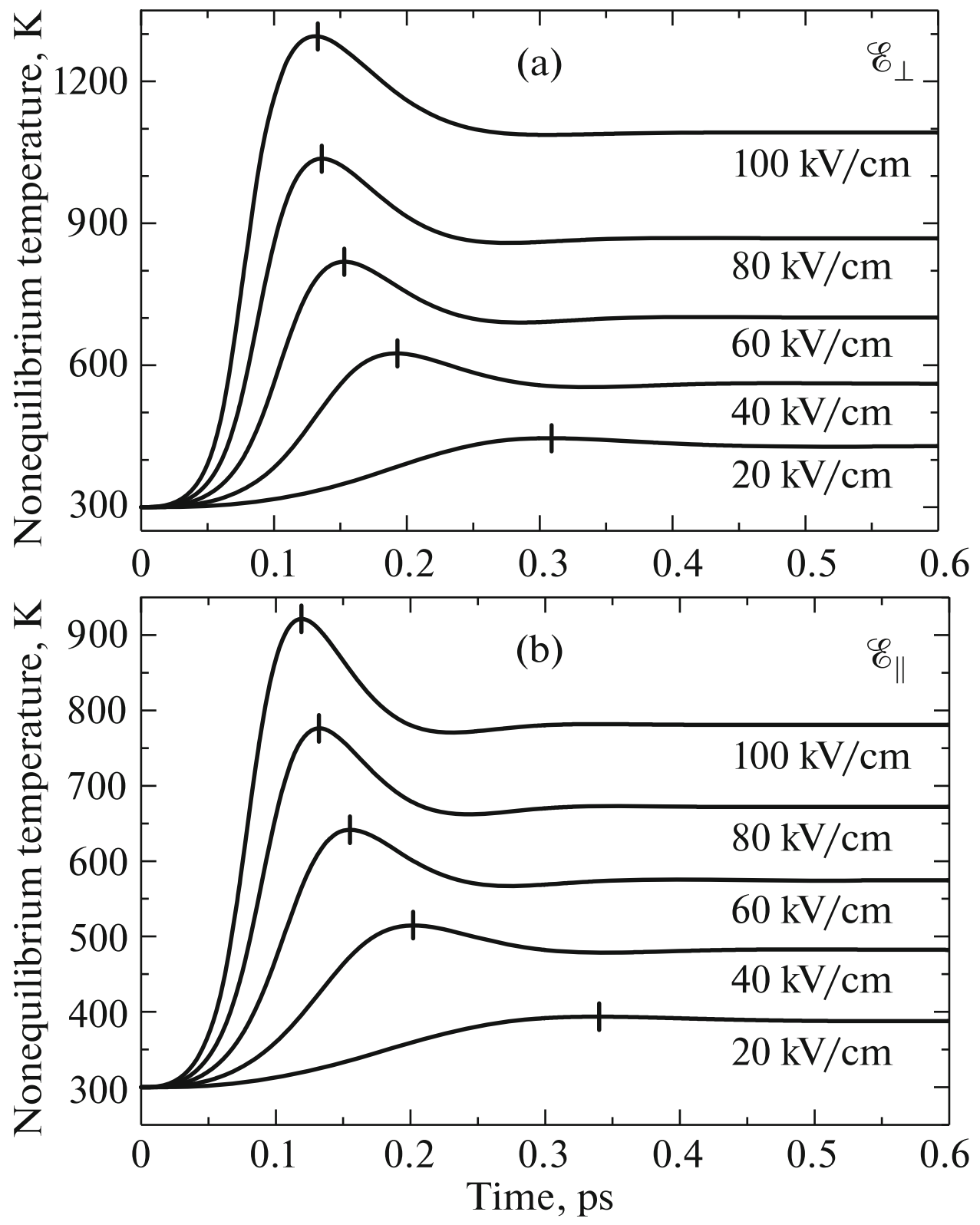

**Fig. 1.** Non-equilibrium temperature of electrons versus time, for $n$-type doped semiconductor 6$H$-SiC: (a) for $\mathscr{E}_\perp$, (b) for $\mathscr{E}_\parallel$.

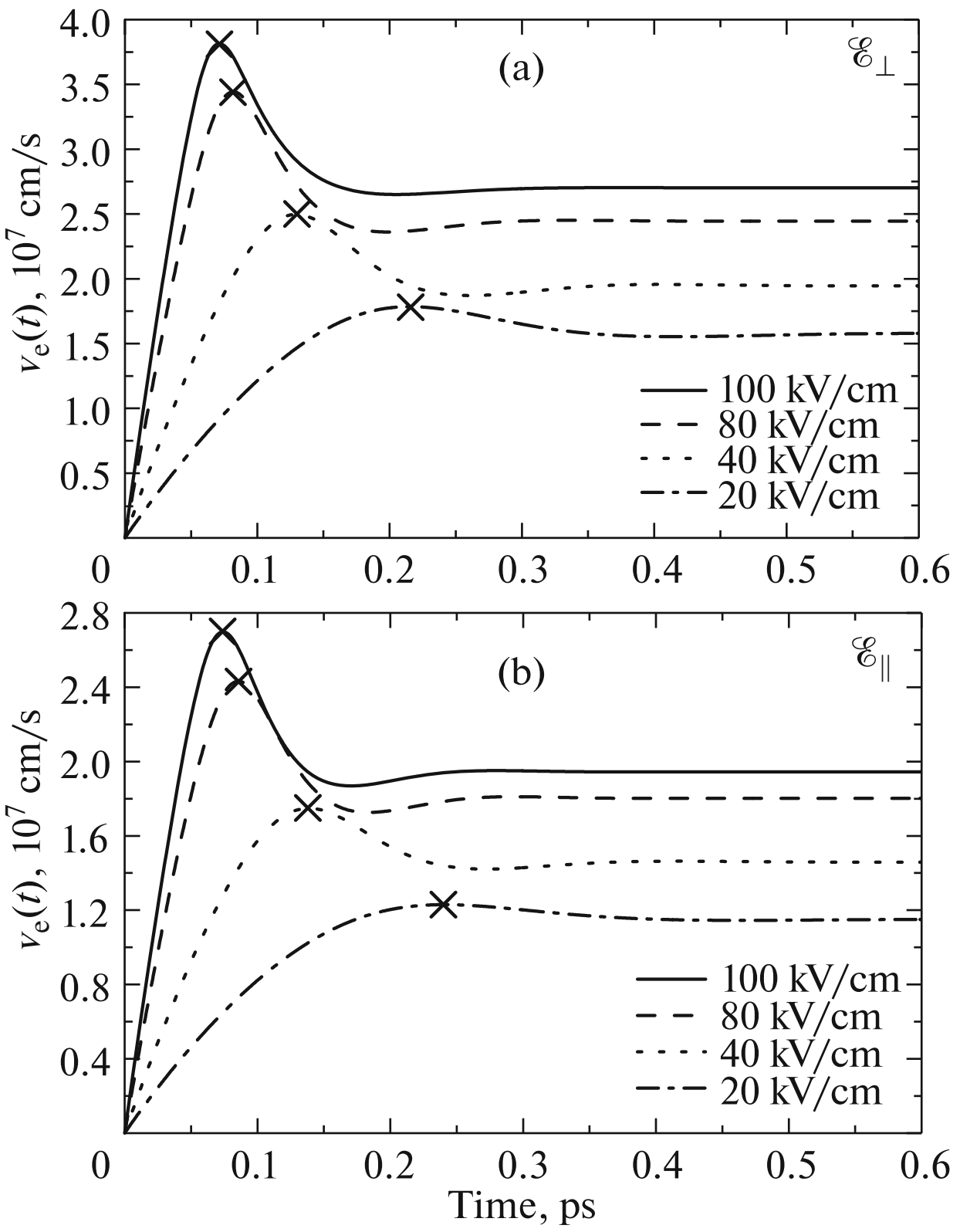

**Fig. 2.** Drift velocity of electrons versus time for $n$-type doped semiconductor 6$H$-SiC: (a) for $\mathscr{E}_\perp$, and (b) $\mathscr{E}_\parallel$.

where

$$\mu_e^*(t) = \frac{1}{\beta_e^*(t)} \ln\left\{ \frac{4n\hbar^3}{e^{\beta_e^*(t) m_e^* v_e(t)^2/2}} \sqrt{\left(\frac{\pi\beta_e^*(t)}{2m_e^*}\right)^3} \right\}. \quad (14)$$

We emphasize that we take the "Einstein model" dispersionless frequency relation for LO phonons and the "Debye model" dispersionless frequency relation for AO phonons.

## 3. RESULTS

By using the Nonequilibrium quantum kinetic theory [22] derived from NSO [6–15] the basic macrovariables of the system and the time evolution are obtained after numerically resolving the set of differential Eqs. (4)–(7) (which are nonlinear equations). We assumed that the reservoir temperature is 300 K, and an electron concentration $n$ equal to the ionized dopant concentration $n_I$, that is, $n = n_I = 10^{16}$ cm$^{-3}$. We have used the parameters for the semiconductor 6$H$-SiC extracted from [27–29], namely: electron effective mass for $\mathscr{E}_\perp$: $m_{e\perp}^* = 0.24m_0$ [27]; electron effective mass for $\mathscr{E}_\parallel$: $m_{e\parallel}^* = 0.34m_0$ [27]; lattice constant $a = 3.080$ Å [28]; lattice constant $c = 15.120$ Å [28]; longitudinal optical phonon energy $\hbar\omega_{LO} = 0.12$ eV [29]; mass density = 3.2 g/cm$^3$ [29]; sound velocity $v_s = 1.355 \times 10^6$ cm/s [29]; acoustic deformation potential $E_1 = 11.2$ eV [29]; low frequency dielectric constant $\epsilon_0 = 9.7$ [29]; high frequency dielectric constant $\epsilon_\infty = 6.5$ [29].

Figures 1 and 2 show, respectively, the nonequilibrium temperature of electrons ($T_e^*$) and drift velocity of electrons ($v_e$) as a function of time (in picoseconds) for various values of the electric field intensity. In Figs. 1a and 2a $\mathscr{E}_\perp$ is concerned, while Figs. 1b and 2b relates to $\mathscr{E}_\parallel$. The time for the electrons to attain the steady state is about 0.5 ps. Figures 1 and 2 show an overshoot in the nonequilibrium temperature of electrons and in the drift velocity of electrons. This phenomenon occurs if the carrier relaxation rate of energy is greater than the carrier relaxation rate of momentum during the evolution dynamics of the macroscopic state of the system.

The vertical marks at the curves of Fig. 1 and the crosses at the curves of Fig. 2 indicate the points where the maximum values for each curve occur ("peak points"). It is possible to notice that these points are shifted towards the $Y$-axis with the increase of the elec-tric fields. We denote by $t_p$ the instant of time at which the peaks occur in each curve. Note that these peaks do not occur at the same instant for nonequilibrium temperature of electrons and for drift velocity of electrons (remembering that the relaxation rates of momentum and energy are different). The times $t_p$ are shown in Fig. 3 for nonequilibrium temperature of electrons and for drift velocity of electrons, for $\mathscr{E}_\perp$ (a) and $\mathscr{E}_\parallel$ (b), indicating evident decrease of the time $t_p$ with the increase of the intensity of the applied electric field.

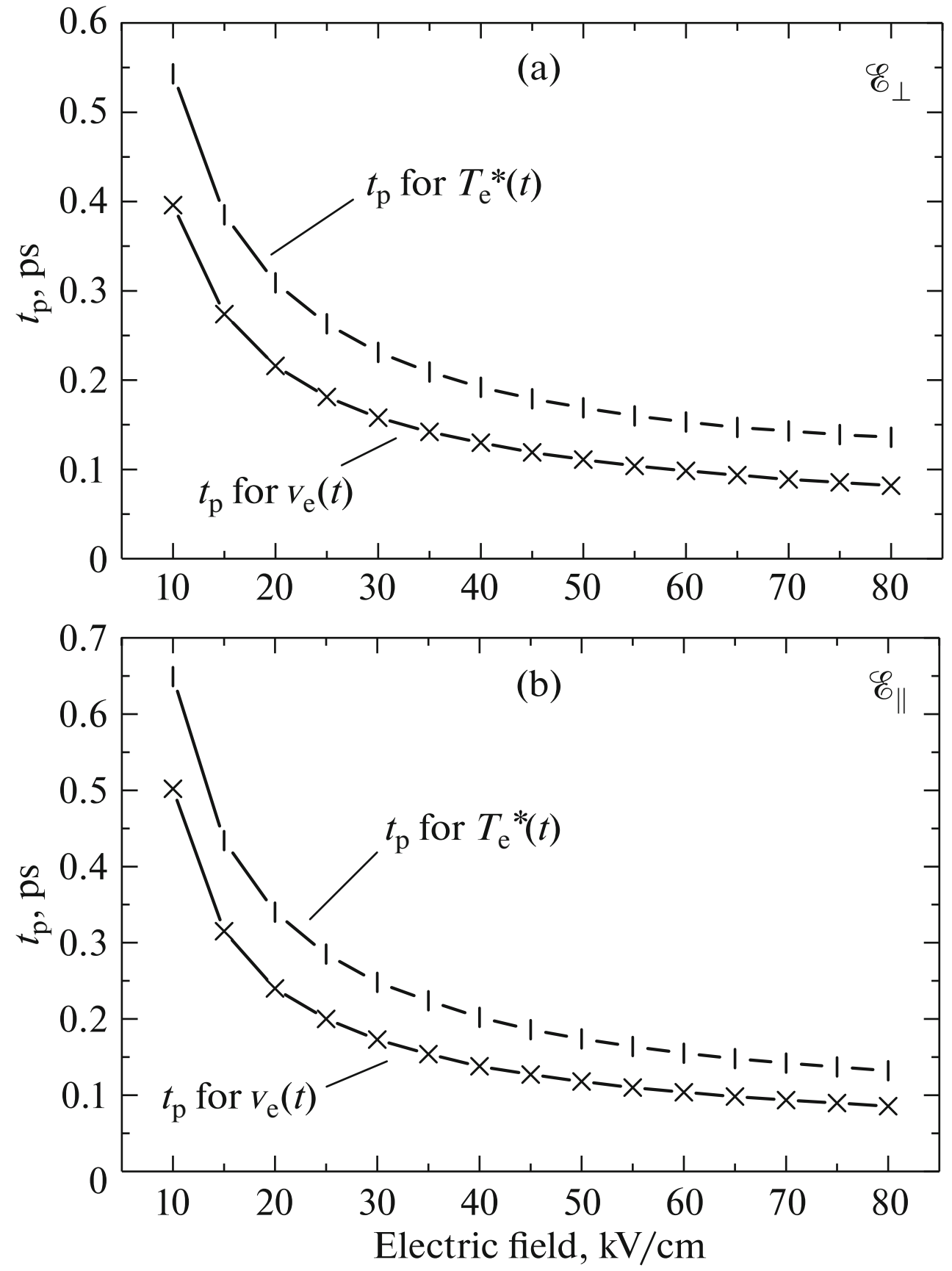

**Fig. 3.** The time $t_p$ versus electric field magnitude for $n$-type doped semiconductor 6$H$-SiC: (a) for $\mathscr{E}_\perp$, and (b) for $\mathscr{E}_\parallel$.

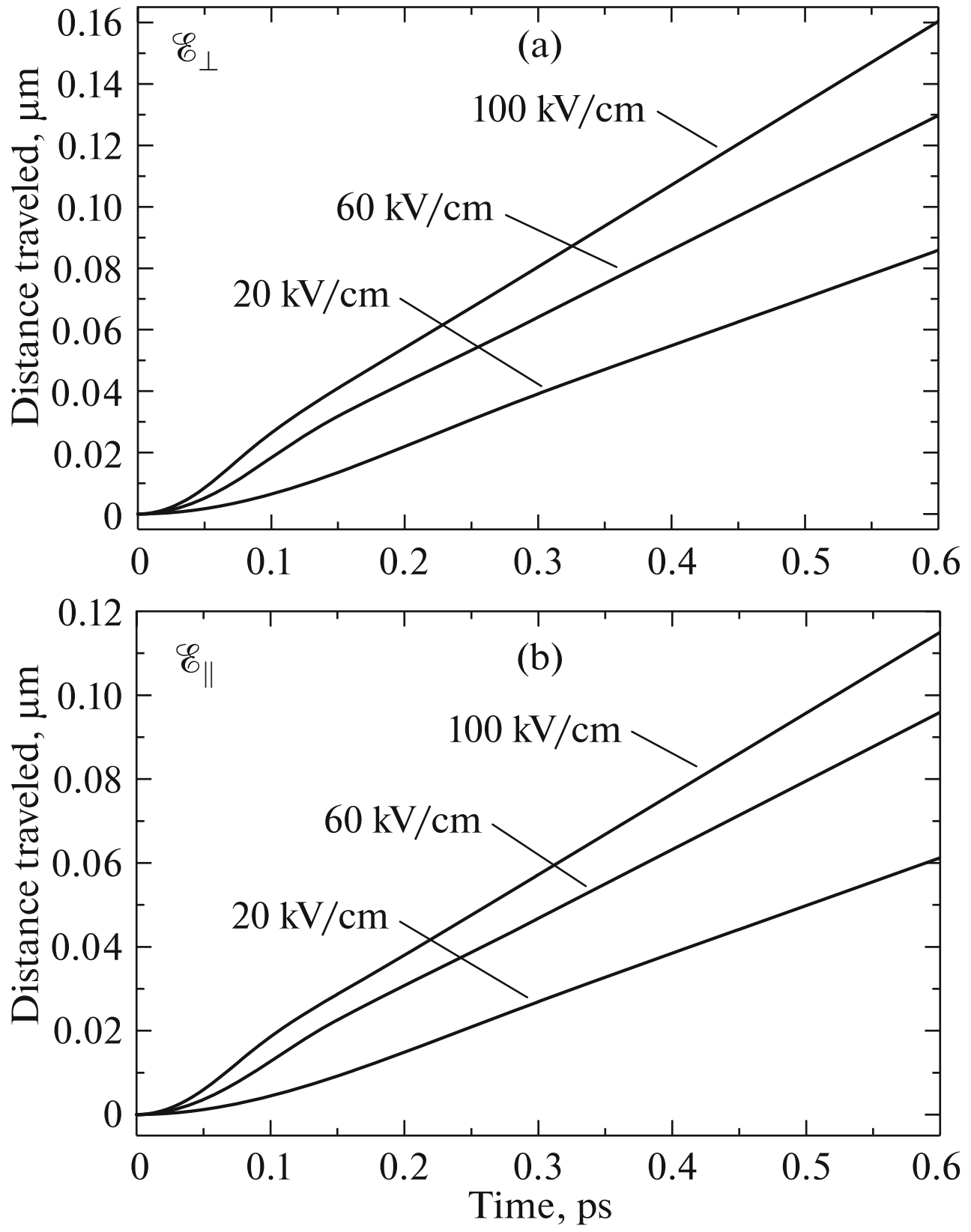

**Fig. 4.** Electron distance traveled versus time in $n$-type doped semiconductor 6$H$-SiC for three values of electric field magnitude: (a) for $\mathscr{E}_\perp$, (b) for $\mathscr{E}_\parallel$.

Figure 4 shows the electron distance traveled in $n$-type doped semiconductor 6$H$-SiC for three values of the electric field intensity. At a time interval of 0.6 ps, the values of the distance traveled are in the order of sub μm, and we note that the electron distance traveled for $\mathscr{E}_\perp$ (Fig. 4a) is approximately 1.4 times the distance traveled for $\mathscr{E}_\parallel$ (Fig. 4b). For the electric field intensity of 100 kV/cm, for example, the electron distance traveled for $\mathscr{E}_\perp$ is approximately 0.16 μm and that for $\mathscr{E}_\parallel$ is approximately 0.11 μm, an absolute difference of 0.05 μm.

Figure 5 shows, in the steady state, the increase of nonequilibrium temperature of electrons (in kelvins) with the increase of applied electric field strength in $n$-type doped semiconductor 6$H$-SiC. The increase in the nonequilibrium temperature of electrons comes from the interaction of the carriers with the applied electric field. The increase in nonequilibrium temperature of the longitudinal optical phonons and in the nonequilibrium temperature of acoustical phonons (not shown here) was insignificant: less than 1%. We emphasize that we take the "Einstein model" dispersionless frequency relation for longitudinal optical phonons and the "Debye model" dispersionless frequency relation for acoustical phonons.

Figures 6 shows the dependence of the drift velocity of electrons, in the steady state, on the electric field magnitude in $n$-type doped semiconductor 6$H$-SiC. In Figs. 5 and 6, the upper curve (solid line) is for $\mathscr{E}_\perp$ and the lower curve (dashed line) is for $\mathscr{E}_\parallel$. Looking at Fig. 6, it can be noticed that the highest velocity corresponds to $\mathscr{E}_\perp$. Of course, this follows from the fact that the electrons have a lower effective mass in the orientation perpendicular to the $c$-axis than in the orientation parallel to the $c$-axis. In addition, it is noted

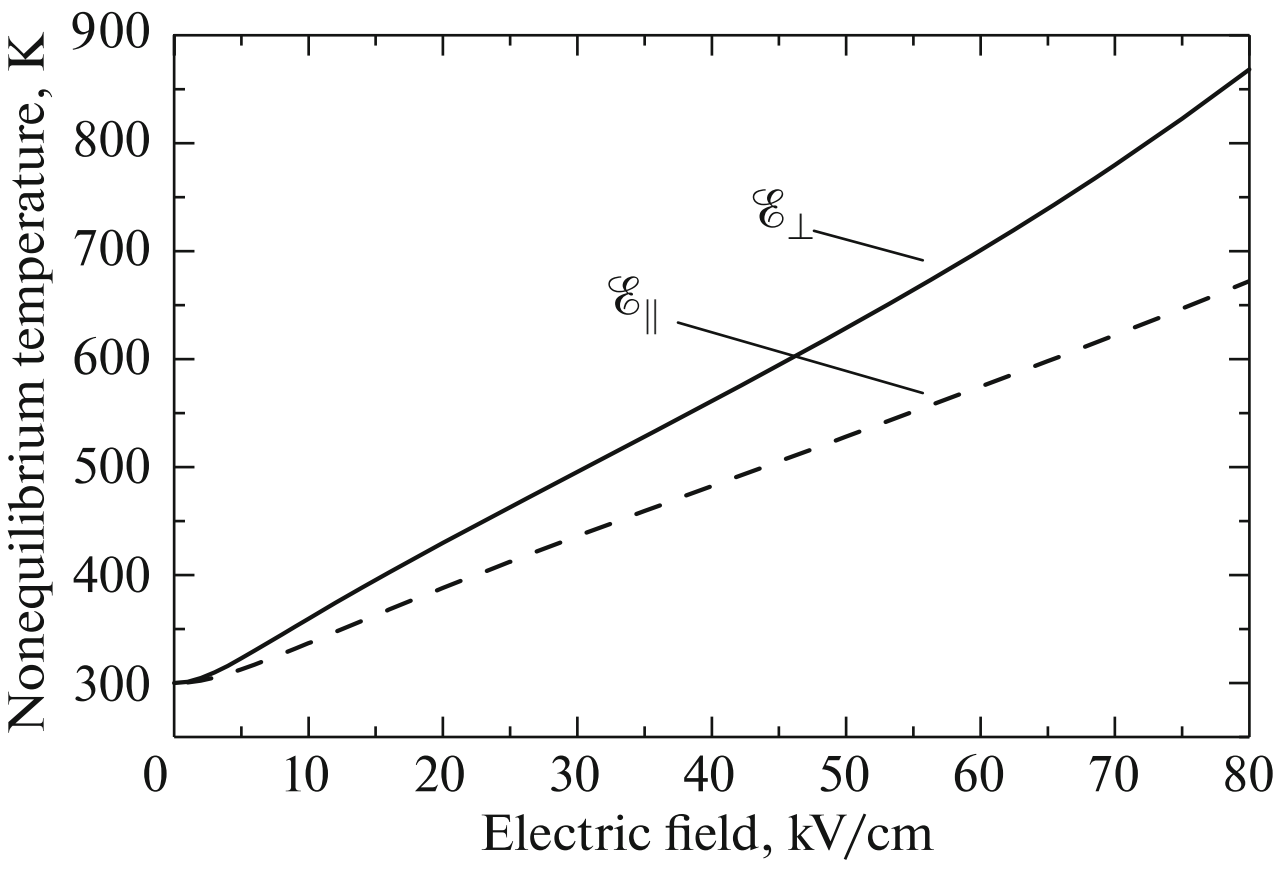

**Fig. 5.** Non-equilibrium temperature of electrons versus electric field magnitude for *n*-type doped semiconductor 6*H*-SiC. Solid line: $\mathscr{E}_\perp$. Dashed line: $\mathscr{E}_\parallel$.

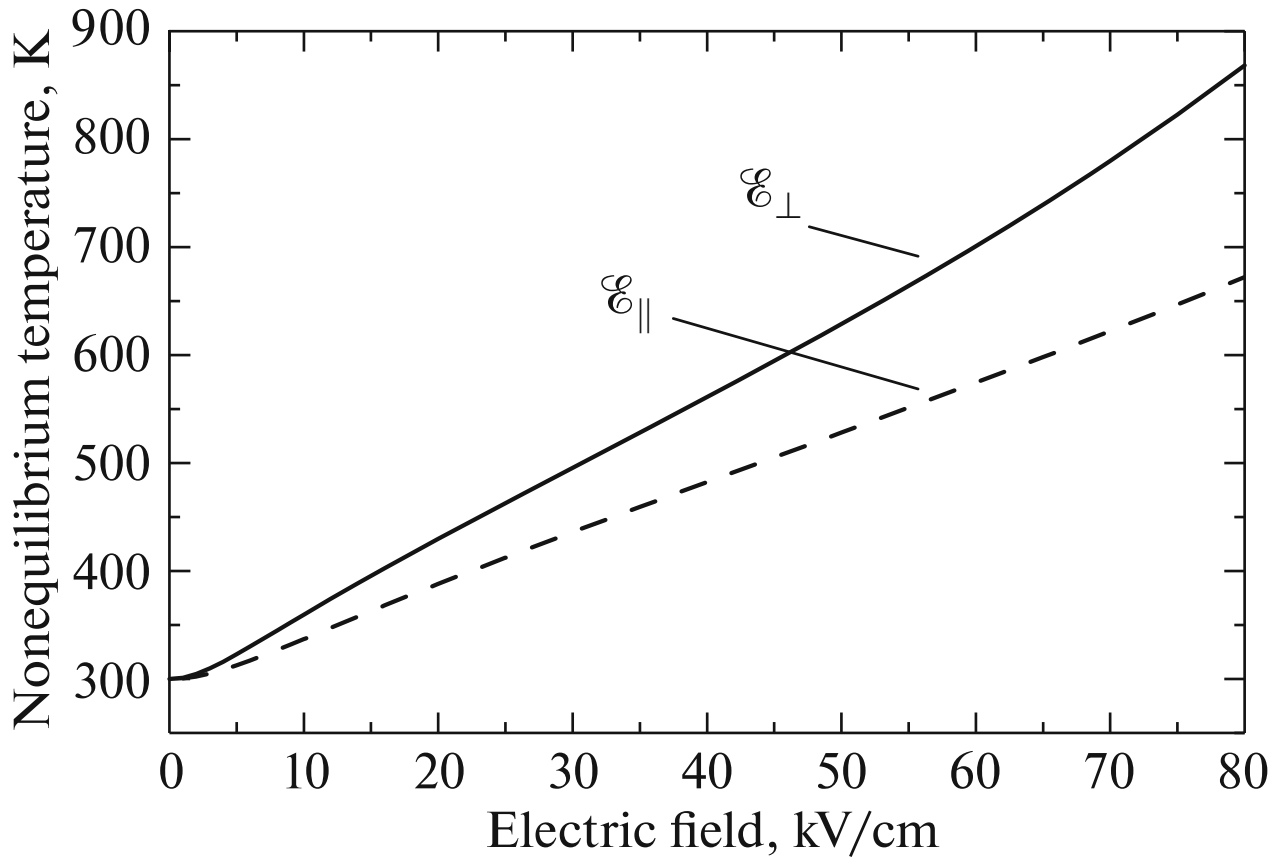

**Fig. 6.** Drift velocity of electrons versus electric field magnitude for *n*-type doped semiconductor 6*H*-SiC. Solid line: $\mathscr{E}_\perp$, dashed line: $\mathscr{E}_\parallel$.

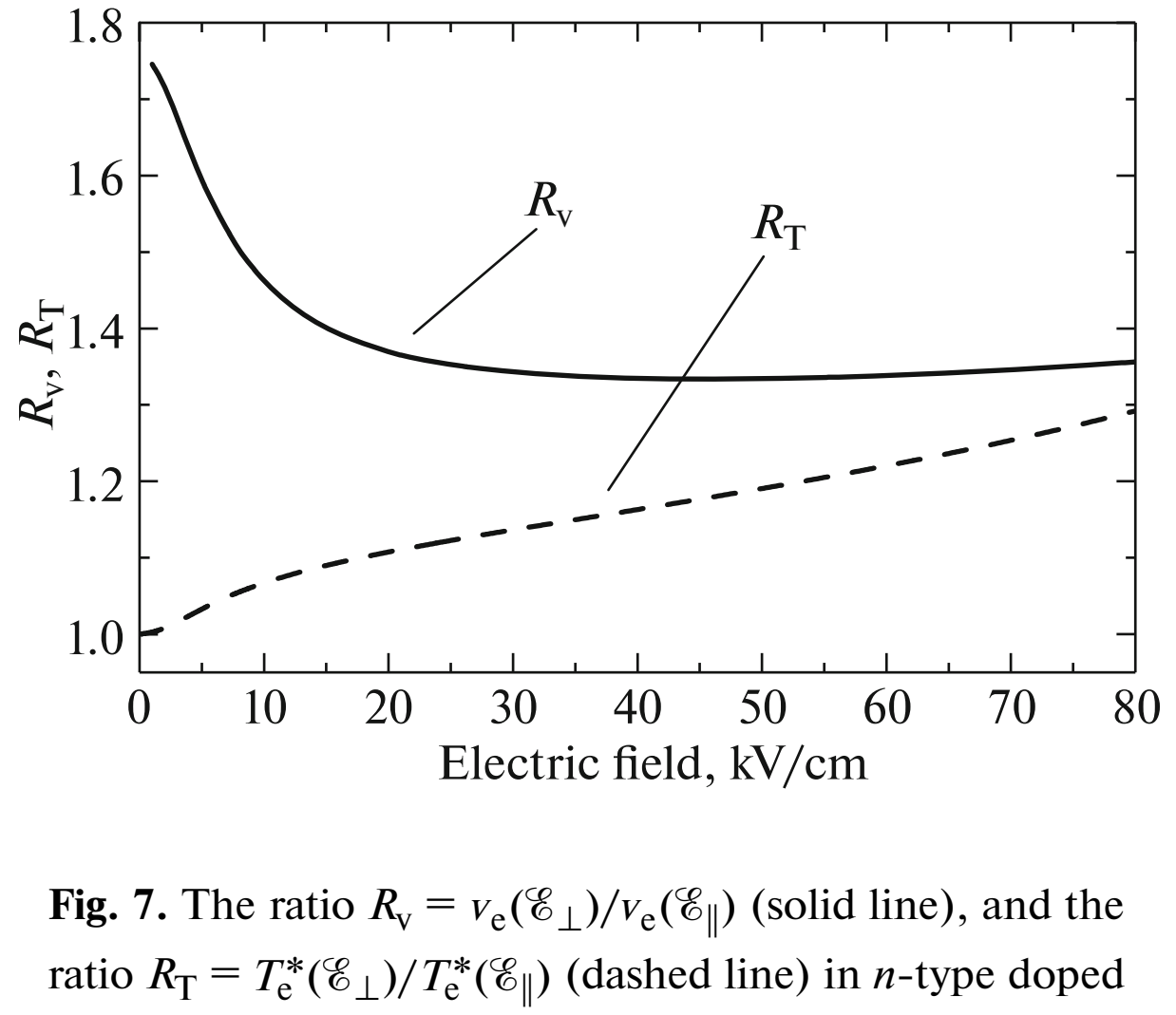

**Fig. 7.** The ratio $R_v = v_e(\mathscr{E}_\perp)/v_e(\mathscr{E}_\parallel)$ (solid line), and the ratio $R_T = T_e^*(\mathscr{E}_\perp)/T_e^*(\mathscr{E}_\parallel)$ (dashed line) in *n*-type doped semiconductor 6*H*-SiC.

that there is an ohmic regime for electric fields less than 2 kV/cm.

In Fig. 7, the dashed line shows the ratio $R_T$ between the nonequilibrium temperature of electrons with the electric field intensity applied perpendicular to the *c*-axis ($\mathscr{E}_\perp$) and the nonequilibrium temperature of electrons with the electric field intensity applied parallel to the *c*-axis ($\mathscr{E}_\parallel$), that is: $R_T = T_e^*(\mathscr{E}_\perp)/T_e^*(\mathscr{E}_\parallel)$. Moreover, the solid line in Fig. 7 shows the ratio $R_v$ between the drift velocity of electrons with the electric field intensity applied perpendicular to the *c*-axis ($\mathscr{E}_\perp$) and the drift velocity of electrons with the electric field intensity applied parallel to the *c*-axis ($\mathscr{E}_\parallel$), that is: $R_v = v_e(\mathscr{E}_\perp)/v_e(\mathscr{E}_\parallel)$. Inspection of Fig. 7 tells us that the difference between $T_e^*(\mathscr{E}_\perp)$ and $T_e^*(\mathscr{E}_\parallel)$ increases with increasing of the magnitude of the electric field applied reaching a value of 29% at 80 kV/cm, while the difference between $v_e(\mathscr{E}_\perp)$ and $v_e(\mathscr{E}_\parallel)$ remains in a range of 33–74%.

## 4. FINAL REMARKS

Summarizing, in this paper we present a study on the charge transport in *n*-type doped 6*H*-SiC (in transient and steady state) using a nonequilibrium quantum kinetic theory derived from the method of nonequilibrium statistical operator (NSO). To develop high performance electronic devices, beyond optimizing the fabrication steps, a good knowledge of the transport properties is required. As an example, the carrier mobility is a very important property, affecting the device performances and, hence, can be considered as a figure of merit of the microscopic quality of the epilayers. Depending on the particular device, one is interested in the carrier mobility either perpendicular to the *c*-axis or parallel to it. In particular, the carrier mobility along the *c*-axis is extremely important in vertical power devices such as Schottky diodes.